\documentclass[aps,prl,twocolumn,superscriptaddress,groupedaddress]{revtex4}  
\usepackage{graphicx}  
\usepackage{dcolumn}   
\usepackage{bm}        
\usepackage{amssymb}   
\usepackage{amsmath}
\usepackage{subfigure}
\usepackage{bbm}

\begin{document}



\title{Mapping quorum sensing onto neural networks to understand collective decision making in heterogeneous microbial communities}
\author{Tahir I. Yusufaly}
\affiliation{Department of Physics and Astronomy, University of Southern California, Los Angeles, CA, 90089}
\author{James Q. Boedicker}
\affiliation{Department of Physics and Astronomy, University of Southern California, Los Angeles, CA, 90089}
\affiliation{Department of Biological Sciences, University of Southern California, Los Angeles, CA, 90089}
\date{\today}

\begin{abstract}

Microbial communities frequently communicate via quorum sensing (QS), where cells produce, secrete, and respond to a threshold level of an autoinducer (AI) molecule, thereby modulating gene expression. However, the biology of QS remains incompletely understood in heterogeneous communities, where variant bacterial strains possess distinct QS systems that produce chemically unique AIs. AI molecules bind to `cognate' receptors, but also to `non-cognate' receptors found in other strains, resulting in inter-strain crosstalk. Understanding these interactions is a prerequisite for deciphering the consequences of crosstalk in real ecosystems, where multiple AIs are regularly present in the same environment. As a step towards this goal, we map crosstalk in a heterogeneous community of variant QS strains onto an artificial neural network model. This formulation allows us to systematically analyze how crosstalk regulates the community's capacity for flexible decision making, as quantified by the Boltzmann entropy of all QS gene expression states of the system. In a mean-field limit of complete cross-inhibition between variant strains, the model is exactly solvable, allowing for an analytical formula for the number of variants that maximize capacity as a function of signal kinetics and activation parameters. An analysis of previous experimental results on the \textit{Staphylococcus aureus} two-component Agr system indicates that the observed combination of variant numbers, gene expression rates and threshold concentrations lies near this critical regime of parameter space where capacity peaks. The results are suggestive of a potential evolutionary driving force for diversification in certain QS systems.

\end{abstract}
\maketitle

Multicellular communities of microbes frequently coordinate changes in group behavior, which requires cell-to-cell communication via the exchange of extracellular signaling molecules called autoinducers (AI), a process known as quorum sensing (QS) \cite{QSReview}. Microbes produce AI molecules at a default basal rate, and if the local concentration of AI molecules accumulates beyond a certain threshold, the AI production rate is amplified to an activated level via a positive-feedback loop, simultaneously regulating the expression of QS-controlled genes. This two-state QS regulatory network controls a wide array of collective behaviors in microbial communities, such as the formation of biofilms, the regulation of virulence, or lateral gene transfer \cite{QSReview, WilliamsReview, JayaramanReview, VegaGore2014, NealsonBoedickerReview}.

\begin{figure}
\subfigure[]{\includegraphics[scale=0.3]{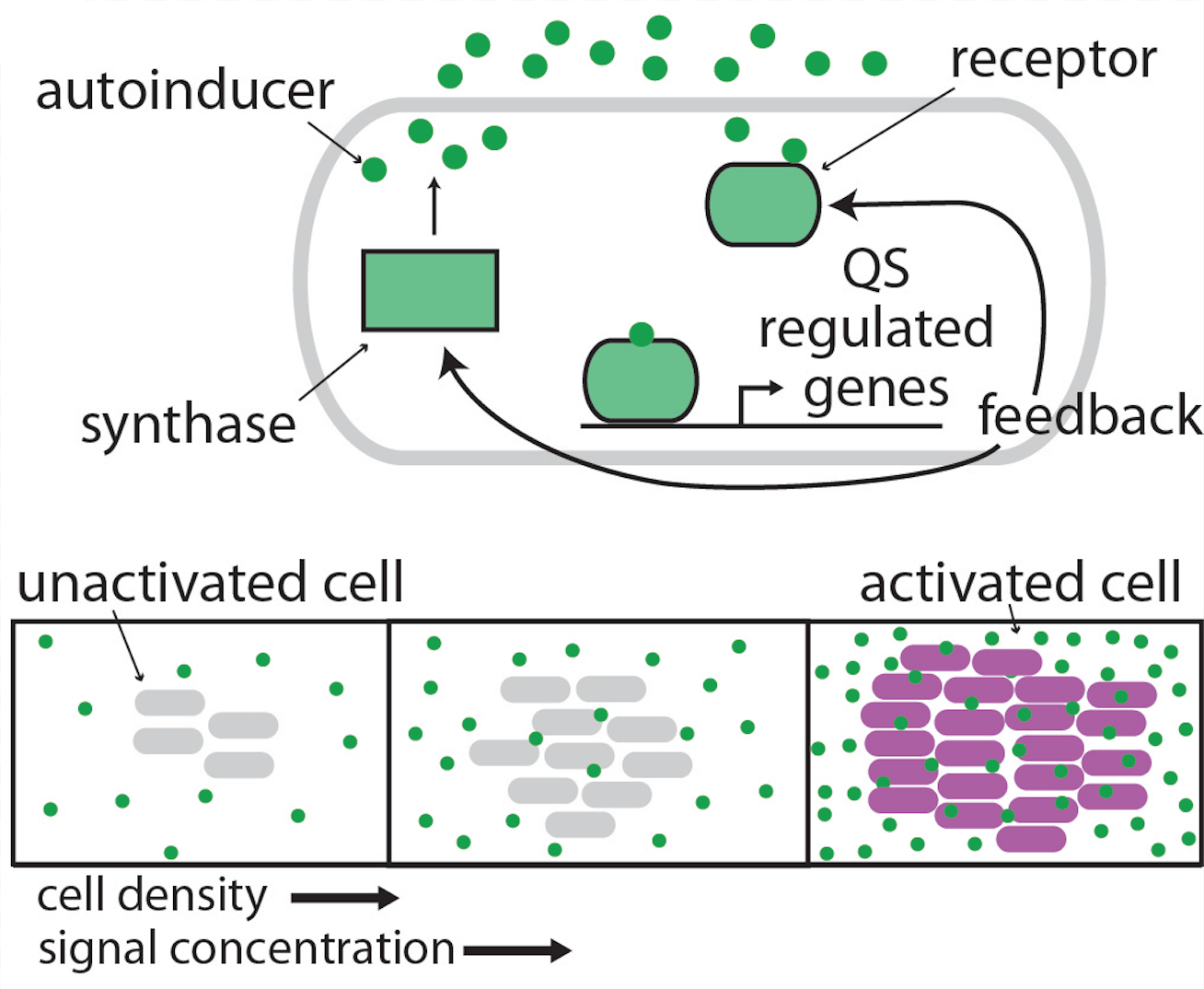}
\label{subfig:1a}}
\subfigure[]{\includegraphics[scale=0.3]{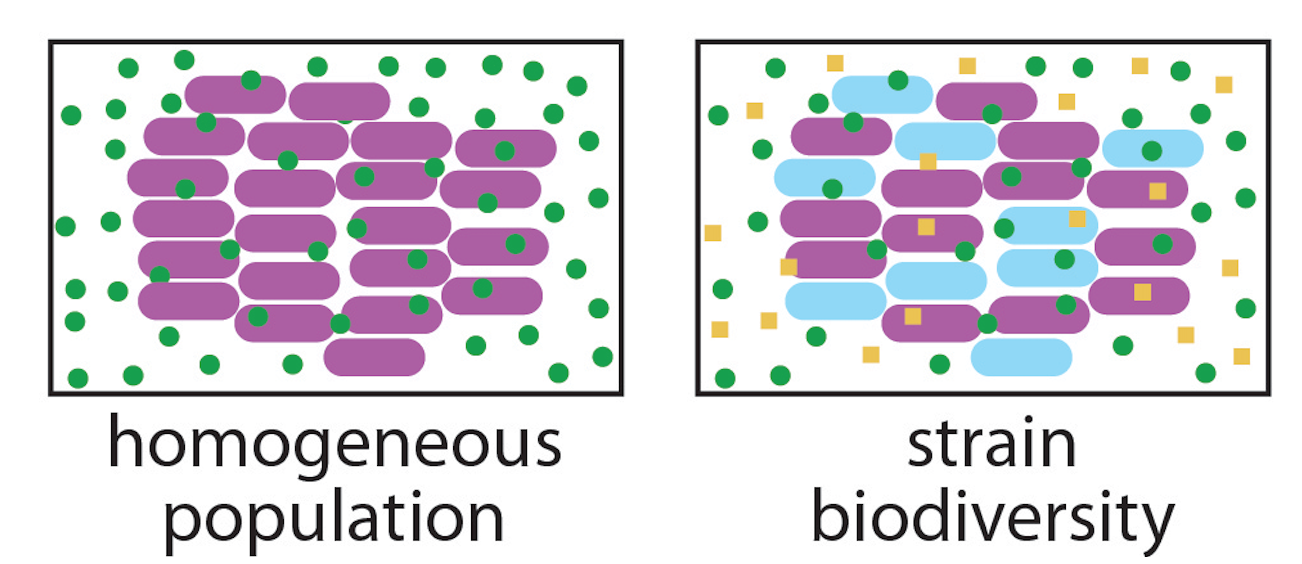}
\label{subfig:1b}}
\caption{(a) Quorum sensing (QS) has traditionally been studied in homogeneous populations, with a single regulatory circuit consisting of a synthase that produces an autoinducer (AI), and a receptor that senses the local concentration of the same AI. At low cell density, the synthase genes are expressed at a low basal level, secreting a small amount of AI into the environment. Once the colonies have grown beyond a critical population density, the collective AI concentration exceeds the threshold required to activate QS. (b) In reality, QS systems are characterized by substantial heterogeneity. A common origin of this is in the coexistence of multiple variants (colored purple and light blue) of the same QS system in the population, each of which produces variant AI molecules (green circles and gold squares, respectively). These variant systems have different specificities of signal reception, and as a result, there can be crosstalk between variant subgroups, leading to novel effects not present in homogeneous populations.}
\label{Fig:SpatiotemporalSignalingDynamics}
\end{figure}

Although quorum sensing has traditionally been viewed as a process associated with homogeneous populations, several results in recent years have called this conventional wisdom into question \cite{PradhanAndChatterjee2014,Grote2015,PerezVelazquez2015}. In particular, it is by now well established that real microbial communities are frequently characterized by the stable coexistence of several variant QS systems in the population \cite{SocialEvolutionPLOS2016,FacCheatingPNAS2016,FEMSReview2016}. AI molecules produced by cells with one QS variant tend to activate QS in related kin cells that also express that same variant, with the corresponding native cognate receptor. However, when multiple variants are simultaneously present in the community, nonnative AI molecules can spuriously bind to non-cognate receptors in different strains \cite{PerezWeissHagen,PavansPaper}. This results in crosstalk between strains - distinct variants of a QS allele interact with and modify each other's signaling behavior \cite{MehtaWingreen}, as visualized in Figure \ref{Fig:SpatiotemporalSignalingDynamics}. In recent years, it has become increasingly evident that this crosstalk amounts to more than just unwanted noise - instead, heterogeneous combinations of interactions between members of a community provide a tunable `knob' controlling the structure and function of the global QS network \cite{WuWang2014,NaturalDiversityofQS2015,ScottHasty2016}.

In light of these developments, it is clear that traditional quantitative models of QS \cite{Dockery2001, Chopp2002}, which ignore crosstalk between different QS systems, are insufficient for explaining the complete diversity of QS behaviors found in nature. In order to understand, predict and control QS in these more complex communities, more sophisticated approaches are needed. 

In this manuscript, we present such an approach. We demonstrate that crosstalk between diverse genetic QS variants can be explicitly mapped onto an artificial neural network. This formalism allows for a much broader investigation of possible signaling behaviors in QS communities. As an explicit example of the utility of our formalism, we consider the scenario of completely inhibitory crosstalk between variant strains, which is equivalent to well-known local-excitation, global-inhibition (LEGI) models of neural coding \cite{Xiong2010, Deco2014}, and calculate the optimal number of distinct strains needed to maximize the capacity of the network, as quantified by the Boltzmann entropy. We find that this number, which lies at a phase boundary, is consistent with patterns of diversification found in the two-component Agr QS system of \textit{Staphylococcus aureus} \cite{QSStaph2003Review}, suggesting a possible selective constraint on the evolution of certain classes of QS networks.

\section{Mathematical Formalism}

\subsection{Single Strain}
To start, consider a single bacterial strain, interacting via only one autoinducer pathway, and well-mixed in a homogeneous medium. Suppose that we are in a steady-state regime, such that the total density of cells in the community, $c_{cells}$, is fixed. Furthermore, suppose that each cell produces autoinducer at a basal rate $r_{b}$, and thus, that $c_{cells} r_{b}$ is the total basal rate of all cells. We also allow the autoinducer to decay with decay constant $\gamma$. In addition, once the total concentration of AI molecules $n$ in the system is above the critical threshold $n_{c}$, we suppose that the total activity of each cell adjusts to an activated level $r_{a} = f r_{b}$ where $f > 1$ is the fold-change. In general, this activation is more accurately described by a smooth graded response, frequently represented as a Hill function \cite{HillActivation}, but as a simplified initial approximation, we can treat it as an instantaneous Heaviside step-function $H$ \cite{StepActivation}. Then, the dynamics of the autoinducer are encoded in the equation:
\begin{equation}
\frac{dn}{dt} = c_{cells} r_{b}(1 + (f-1) \ H(n - n_{c})) - \gamma n .
\label{eq:ProductionDecayActivation}
\end{equation}

\subsection{Multiple Strains in the Mean Field Approximation}
Now, consider the case where the original strain differentiates into multiple distinct strains.  For simplicity, we will make a `mean-field' approximation to all parameters - in other words, we will assume that we can replace each parameter with its average value for all the different strains in the system. Thus, for example, if there are two variant strains, producing signal $n_{1}$ and $n_{2}$ with rates $r_{b1}$ and $r_{b2}$, respectively, we take the approximation that $r_{b1} \approx r_{b2} \approx  r_{b} $, which we take to be the average basal rate of all the cells in the system. We can likewise take the fold change $f$, threshold $n_{c}$, decay rate $\gamma$, and number of cells per strain to all be equal to their average values. 

For concreteness, let us now focus specifically on the case of two variants. In the mean-field approximation, then, both strains are equally likely, and so the original total basal rate $c_{cells} r_{b}$ of all the cells now becomes split into a new total $c_{cells} r_{b}/2$ per strain. In addition, we must now account for crosstalk between strains. Specifically, although the receptors of strain 1 will likely bind most strongly to their cognate signal $n_{1}$, they also have a non-zero affinity for signal $n_{2}$, and vice versa for the receptors of strain 2 binding to signal $n_{1}$. We can represent these additional affinities by generating `effective' concentrations felt by strains 1 and 2,
\begin{subequations}
\begin{align}
n^{eff}_{1} = n_{1} + w_{12} n_{2} \\
n^{eff}_{2} = w_{21} n_{1} + n_{2} 
\label{eq:effective}
\end{align} 
\end{subequations}
where $w_{12}$ is a weight term that quantifies the relative affinity of strain 1's receptors for signal $n_{2}$ compared to $n_{1}$, and likewise for $w_{21}$. 

With this notation, we can now generalize the autoinducer dynamics to:
\begin{subequations}
\begin{align}
\frac{dn_{1}}{dt} = \frac{c_{cells} r_{b}}{2}(1 + (f-1) \ H(n^{eff}_{1} - n_{c})) - \gamma n_{1} \\
\frac{dn_{2}}{dt} = \frac{c_{cells} r_{b}}{2}(1 + (f-1) \ H(n^{eff}_{2} - n_{c})) - \gamma n_{2} .
\end{align} 
\end{subequations}

The physical interpretation of the weights is straightforward: for instance, a smaller $w_{12}$ means that each individual $n_{2}$ molecules has little affinity for strain 1, and thus, it takes a correspondingly greater number of $n_{2}$ molecules to generate the same effect of a single $n_{1}$ molecules. Furthermore, if $w_{12} < 0$, that implies that $n_{2}$ has an inhibitory effect on strain 1, increasing the number of $n_{1}$ molecules required for activation. Note that the weight terms are not necessarily symmetric (i.e., it is possible for $w_{12} \neq w_{21}$). Further microscopic justification for the weights is presented in the Appendix, where it is shown that the weights are approximated by

\begin{equation}
w_{12} = \pm \frac{k_{12}}{k_{11}},
\end{equation}
where $k_{11}$ is the affinity of $n_{1}$ for its cognate receptor $R_{1}$, $k_{12}$ is the affinity of $n_{2}$ for $R_{1}$, and the sign of the weight is positive or negative if the interactions are activating or inhibiting, respectively. In the mean-field-approximation, we can assume all affinities are equal to an average value, $k_{11} \approx k_{12} \approx k_{21} \approx k_{22} \approx k$, in which case all weights are approximately +1 for activation or -1 for inhibition.

Generalizing to $N$ strains $i = 1, 2, ..., N$, we end up with a system of $N$ differential equations
\begin{align}
\frac{dn_{1}}{dt} &= \frac{c_{cells} r_{b}}{N}(1 + (f-1) \ H(n^{eff}_{1} - n_{c})) - \gamma n_{1} \\
& \vdots \nonumber \\ 
\frac{dn_{N}}{dt} &= \frac{c_{cells} r_{b}}{N}(1 + (f-1) \ H(n^{eff}_{N} - n_{c})) - \gamma n_{N} \nonumber 
\end{align}

where
\begin{align}
n^{eff}_{1} &= n_{1} + \dots + w_{1N} n_{N} \\
& \vdots \nonumber \\
n^{eff}_{N} &= w_{N1} n_{1} + \dots + n_{N} \nonumber .
\end{align} 

We once again remind the reader that in the mean-field-limit, all weights $w_{ij} = \pm 1$.

Finally, note that the number of independent parameters may be slightly reduced if we rescale our units into a non-dimensionalised form, 
\begin{equation}
\tilde{n}_{i} = \frac{n_{i}}{c_{cells} r_{b}/\gamma} .
\end{equation}
With this change of variables, we will find the critical AI concentration $n_{c}$ is more conveniently expressed in terms of the corresponding threshold cell density $c_{thresh}$. We can see this by noting that at the threshold, the total basal AI concentration produced by all cells equals the critical concentration,
\begin{equation}
n_{c} = \frac{c_{thresh} r_{b}}{\gamma} .
\end{equation}
Then, non-dimensionalising $n_{c}$, we find
\begin{equation}
\tilde{n}_{c} = \frac{n_{c}}{c_{cells} r_{b}/\gamma} = \frac{c_{thresh}}{c_{cells}} = c_{frac_{crit}}.
\end{equation}
In other words, non-dimensionalisation transforms $n_{c}$ into $c_{frac_{crit}}$, the minimal fraction of the total cell density required for activation (thus, for example, if QS is activated at cell density $c_{cells}/2$, then $c_{frac_{crit}} = 0.5$).

As a result, we obtain the modified set of equations 
\begin{align}
\frac{1}{\gamma} \frac{d\tilde{n}_{1}}{d\tilde{t}} &= \frac{1}{N}(1 + (f-1) H(\tilde{n}^{eff}_{1} - c_{frac_{crit}})) - \tilde{n}_{1} \\
& \vdots \nonumber \\
\frac{1}{\gamma} \frac{d\tilde{n}_{N}}{d\tilde{t}} &= \frac{1}{N}(1 + (f-1) \ H(\tilde{n}^{eff}_{N} - c_{frac_{crit}})) - \tilde{n}_{N} \nonumber
\end{align}

For simplicity, from hereon, we will drop the tildes, setting concentration units such that $c_{cells} r_{b} / \gamma = 1$. 

If we generalize the weights function to allow for self-interaction terms, $w_{ii} = 1$, the above equations can be compressed into a more succinct form
\begin{equation}
\frac{1}{\gamma} \frac{dn_{i}}{dt} = \frac{1}{N}(1 + (f-1) \ H(\sum^{N}_{j=1} w_{ij} n_{j} - c_{frac_{crit}})) - n_{i}
\label{eq:compressed}
\end{equation}
for $i = 1, 2, ..., N$.

\section{Attractor Landscape Analysis}
We can start gaining further insight into the behavior of Equation (\ref{eq:compressed}) by identifying fixed points, or sets of concentrations such that the net production rate equals the net decay rate (i.e, $\frac{dn_{i}}{dt} = 0$ for all $i$). First of all, we note that, since the total (scaled) production rate of a strain can only be one of two values, $1/N$ or $f/N$, setting the production rate equal to the decay rate yields only two possible fixed point concentrations, $n_{i} = 1/N$ or $n_{i} = f/N$. Note however that these are only possible fixed point concentrations - whether or not a given set of ($n_{1}, n_{2}, ..., n_{N}$) concentrations is a fixed point will depend in general on the values of the weight parameters.

To check for whether or not a given set of concentrations is a fixed point, let us define a spin variable $s_{i}$ that can take two values, $s_{i} = 0$ when strain $i$ is inactive, or only producing at a basal level, or alternatively $s_{i} = 1$ when the strain is active, producing at an activated level. Then, the different potential fixed point concentrations are 
\begin{equation}
n_{i} = \frac{1}{N}(1+(f-1)s_{i}) .
\end{equation}
From this setup, it is straightforward to see that in order for a given set of ($s_{1}, s_{2}, ..., s_{N}$) activity levels to be a fixed point, it must satisfy the self consistent condition:
\begin{align} \label{SelfConsistentEquations}
s^{*}_{i} &= H(\frac{f-1}{N} \sum^{N}_{j=1} w_{ij} s^{*}_{j} - c_{frac_{crit}} + \frac{1}{N} \sum^{N}_{j=1} w_{ij} ) \\
	     &= H(\sum^{N}_{j=1} w_{ij} s^{*}_{j} - (\frac{c_{frac_{crit}} - \langle w \rangle_{i}}{(f-1)/N})) \nonumber
\end{align}

where $ \langle w \rangle_{i}$ is the average of the elements in the $i$th row of the weights matrix.
\subsection{Stability of Fixed Points}

We note that any stable attractor set $\vec{s}^{*} = (s^{*}_{1}, s^{*}_{2}, ..., s^{*}_{N})$ which satisfies the self-consistent conditions in Eqs. (\ref{SelfConsistentEquations}) will always be stable. To verify this, recall that if a dynamical system of a vector $\vec{x} = (x_{1}, x_{2}, ... x_{N})$ is defined by
\begin{equation}
\frac{d \vec{x}}{dt} = \vec{f}(\vec{x}) = (f_{1}(\vec{x}), ... f_{N}(\vec{x})),
\end{equation}
then the stability of fixed points is encoded in the behavior of the Jacobian
\begin{equation}
\overleftrightarrow{J} = \frac{d \vec{f}}{d \vec{x}} = (\frac{d \vec{f}}{dx_{1}} ... \frac{d \vec{f}}{dx_{N}}) = \begin{pmatrix}
  \frac{df_{1}}{dx_{1}} & \cdots & \frac{df_{1}}{dx_{N}} \\
  \vdots  & \vdots  & \ddots \\
  \frac{df_{N}}{dx_{1}} & \cdots & \frac{df_{N}}{dx_{N}} 
 \end{pmatrix} .
\end{equation}

Specifically, if $\vec{x}_{0}$ is a fixed point of the system, $\vec{f}(\vec{x}_{0})$ = 0, then the stability of $\vec{x}_{0}$ is determined by following test: if all the eigenvalues of the $\overleftrightarrow{J}(\vec{x}_{0})$ have negative real part, then $\vec{x}_{0}$ is stable, while if even one eigenvalue has positive real part, $\vec{x}_{0}$ is unstable (if the largest eigenvalue is 0, the Jacobian test is inconclusive, but this case is irrelevant for our purposes). 

With this is mind, we note that, as long as we are not at a `crossover' point where $n^{i}_{eff} = c_{frac_{crit}}$ for some $i$, then the Jacobian of the system (\ref{eq:compressed}) is trivially an $N \times N$ matrix of $-1$ along the diagonal, a matrix for which it is obvious by inspection that all eigenvalues must be negative. Furthermore, note that if $n^{i}_{eff} = c_{frac_{crit}}$, then the Heaviside function is undefined and not equal to either 0 or 1, which means that it is impossible for any such point to satisfy (\ref{eq:compressed}) and be a fixed point. Thus, by default, all fixed points are stable attractor states.

\begin{figure}
\subfigure[]{\includegraphics[scale=0.4]{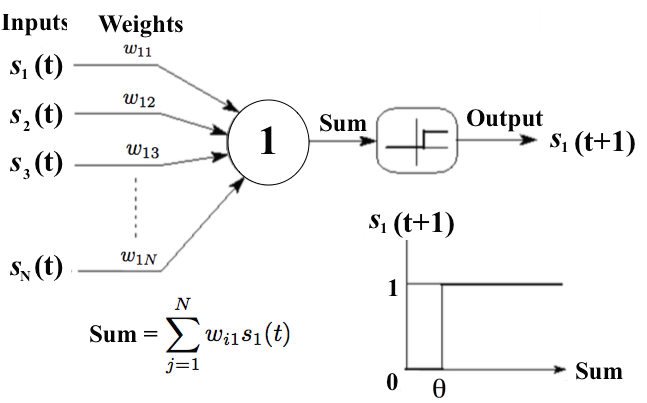}
\label{subfig:2a}}
\subfigure[]{\includegraphics[scale=0.7]{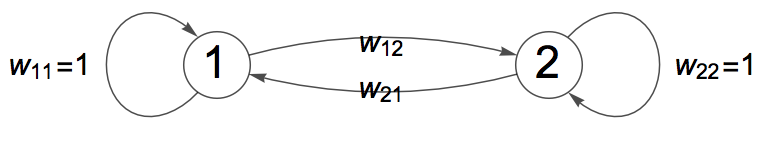}
\label{subfig:2b}}
\caption{A community of quorum-sensing strains can be modeled as an artificial neural network, for example, the Hopfield network of binary McCulloch-Pitts neurons. A single neuron (a) corresponds to a specific strain that produces a chemically distinct AI output, while responding to all the different AI molecules in the environment as input. A population of $N$ distinct strains (b), then, is equivalent to a fully connected network of $N$ neurons, as illustrated for the example of $N = 2$.}
\label{Fig:NNAnalogy}
\end{figure}

\subsection{Connection to Hopfield Network}
In this section we will explicitly highlight the connection between Equation (\ref{SelfConsistentEquations}) and the discrete-time dynamics of a Hopfield network \cite{Hopfield} of binary McCulloch-Pitts \cite{MccullochPitts} neurons. Suppose we have a set of $N$ neurons, with each neuron having one of two possible firing rates, represented by $s_{i} = 0$ (low) or $1$ (high), respectively. If the neurons are fully connected with an $N \times N$ synaptic weight matrix $w_{ij}$, then each neuron activates or suppresses its firing according to the following rule.
\begin{equation}
s_{i}(t+1) = H(\sum^{N}_{j=1} w_{ij} s_{j}(t) - \theta) .
\end{equation}
Here, $s_{i}(t)$ is the activity level of the $i$th neuron at time $t$, $s_{i}(t+1)$ is the updated activity level after one discrete unit of time, and $\theta$ is the threshold synaptic input for activation of a neuron, as shown in Figure \ref{Fig:NNAnalogy}. It is clear that if we solve for the fixed point attractors of the Hopfield network, $s_{i}(t) = s_{i}(t+1) = s^{*}_{i}$, we get precisely the result of Equation (\ref{SelfConsistentEquations}) if we assign $\theta = \frac{c_{frac_{crit}} - \langle w \rangle_{i}}{(f-1)/N}$.

\begin{figure}[t!]
\subfigure[]{\includegraphics[scale=0.24]{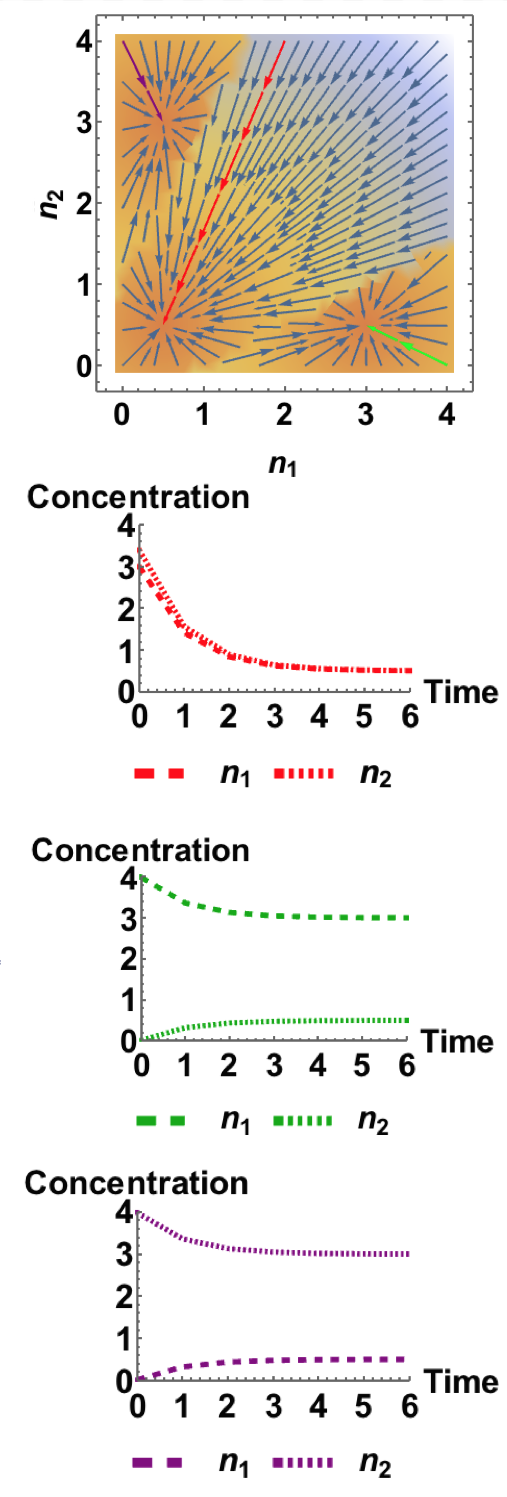}
\label{subfig:3a}}
\subfigure[]{\includegraphics[scale=0.24]{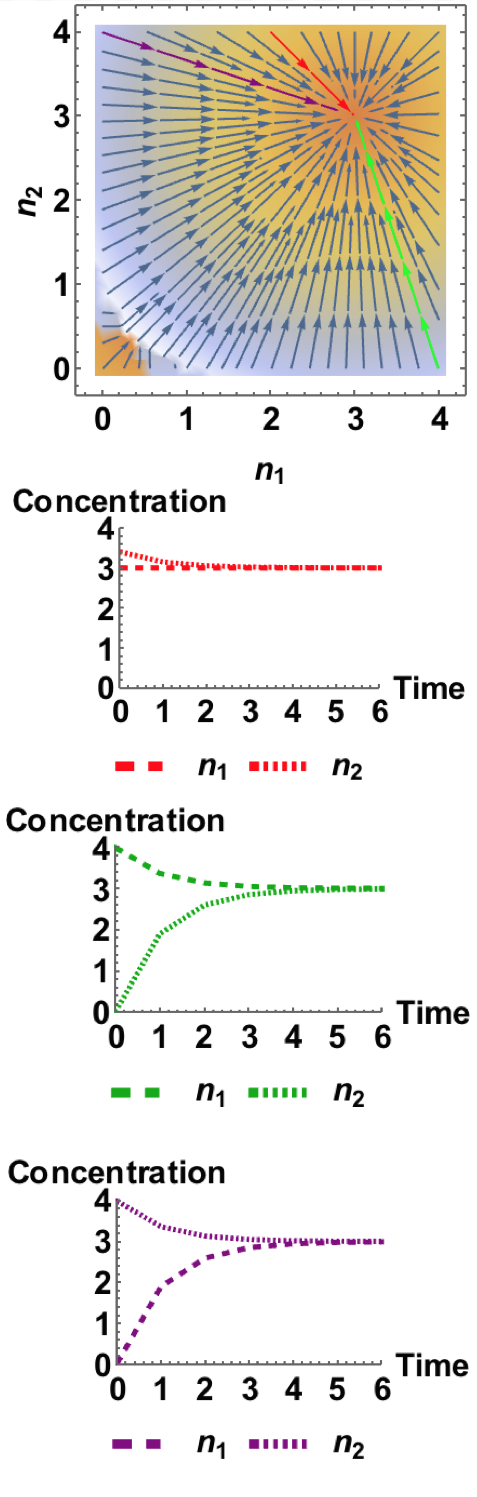}
\label{subfig:3b}}
\caption{Displayed are some example phase portraits of the signaling dynamics of a two-strain system producing autoinducers $n_{1}$ and $n_{2}$ (top row), which can be mapped onto a two-neuron artificial neural network. Each portrait is supplemented with some representative dynamical trajectories of each `neuron' for various initial conditions (bottom three rows). This information is contrasted for two different sets of parameter values. In both instances, the fold change $f$ = 6 and the threshold density $n_{c}$ =  $c_{frac_{crit}}$ = 1 (in non-dimensionalized units). However, the topology of the dynamical system changes depending on the value of the off-diagonal interaction weights, $w_{off-diag}$ = $w_{ij}$ ($i \neq j$). In (a), $w_{off-diag} = -1$, corresponding to cross-inhibition between subgroups. The inhibitory interactions stabilize multiple attractor basins, resulting in the steady-state being dependent on initial conditions. The capacity of this network, as quantified by the entropy of the landscape is Log(3). In (b), $w_{off-diag} = 1$, corresponding to cross-activation between subgroups. This activation essentially guarantees that all systems will activate, resulting in only one stable attractor basin, and the final steady-state being independent of initial conditions. The corresponding neural network landscape has entropy Log(1) = 0, and thus low storage capacity. }
\label{Fig:Landscape}
\end{figure}

\subsection{Measuring Capacity Via Entropy}
So with this setup, we know that for $N$ strains, there are $2^{N}$ possible attractor states $\vec{s}$, of which a certain subset $i = 1, 2, ..., N_{attractor}$ are actual stable attractors $\vec{s}^{*}_{i} = (s^{*}_{1i},...,s^{*}_{Ni})$. Thus, for example, for $N$ = 2, the stable attractors are a subset of four possible choices, $\left \{ \vec{s}^{*} \right \} \subset \left \{ \vec{s} \right \} = \left \{ (0,0), (1,0), (0,1), (1,1) \right \} $.

We may quantify the decision-making capacity \cite{CapacityHopfield} of the community by the Boltzmann entropy \cite{BialekGeneric}:
\begin{equation}
S = \text{log} \ N_{attractor} 
\end{equation}
Intuitively, this makes sense - the larger the number of attractors, the more different decisions the community can make about which cells should or should not activate QS. These ideas are illustrated graphically in Figure \ref{Fig:Landscape}.
 
With these definitions in hand, we may now proceed to look at the behavior of these properties for a particularly tractable model case.

\section{Equal-Strength Cross-Inhibition}
A particularly simple scenario to start with is to just assume that any `non-native', off-diagonal interactions are inhibitory with an antagonistic strength equal and opposite in strength to activation by cognate signals, $w_{ij} = -1$ for $i \neq j$. This maps on to the well-known local-excitation, global-inhibition (LEGI) neural networks, which have been shown to describe how hippocampal neurons represent different associative memories via `cognitive mapping' of distinct memories to distinct activity patterns of place and grid cells \cite{PathIntegration}.

In the case of just one strain, we can almost trivially observe that if the total basal effective concentration of signal (in nondimensionalized units) is 1, the activated effective concentration is $f > 1$, and the threshold cell density fraction is $c_{frac_{crit}} > 0$, then the set of stable attractors is
\begin{equation}
\left \{ s^{*} \right \} = 
  \begin{cases} 
   \left \{ 0 \right \} & \text{if } f \leq c_{frac_{crit}} \\
   \left \{0, 1 \right \}       & \text{if } 1 < c_{frac_{crit}} < f \\
   \left \{1 \right \}       & \text{if } c_{frac_{crit}} \leq 1 \\
  \end{cases}
\end{equation}

For more than one strain, things become less trivial. To start, note that, for multiple strains $N > 1$, the `completely off' state of $N$ zeros, $\vec{s}_{0} = (0,...,0)$, automatically satisfies Eqs. (\ref{SelfConsistentEquations}), and so is always a trivially stable solution. Additionally, having more than one strain on is never stable. To see this, suppose for example that an $N = 2$ strain system were in the state $s_{1} = s_{2} = 1$, implying that $n_{1} = n_{2} = f/2$. However, this means that 

\begin{align}\label{solvedtwostraincase}
n^{eff}_{1} &= w_{11} n_{1} + w_{12} n_{2} \\
&= n_{1} - n{2} \nonumber \\ 
&= 0 < c_{frac_{crit}} \nonumber ,
\end{align}

which by Eqs. (\ref{SelfConsistentEquations}) implies $s_{1} = 0$, a contradiction. It is straightforward to see that the arguments are generalizable to arbitrary numbers of multiple strains.

Finally, let us consider the case where one and only one strain is on - without loss of generality, we can choose $s_{1} = 1$ and all other $s_{2} = ... = s_{N} = 0$. Then, if we require 
\begin{align}\label{solvedoneandonlyone}
n^{eff}_{1} &= w_{11} n_{1} + w_{12} n_{2} + ... + w_{1N} n_{N} \\
&= n_{1} - n_{2} - ... - n_{N} \nonumber \\
&= f/N - 1/N - ... - 1/N  \nonumber \\
&= -1 + (f + 1)/N > c_{frac_{crit}} \nonumber ,
\end{align}
this implies that the state is a stable fixed point only if $N < (f+1)/(c_{frac_{crit}}+1)$.

Thus, summarizing we see that the entropy takes on a simple form,
\begin{equation}
S(N) = 
  \begin{cases} 
   \text{log} \ (N+1)       & \text{if } N \leq (f+1)/(c_{frac_{crit}}+1) .\\
   0       & \text{else } 
  \end{cases}
\end{equation}

\begin{figure}
\includegraphics[scale=.32]{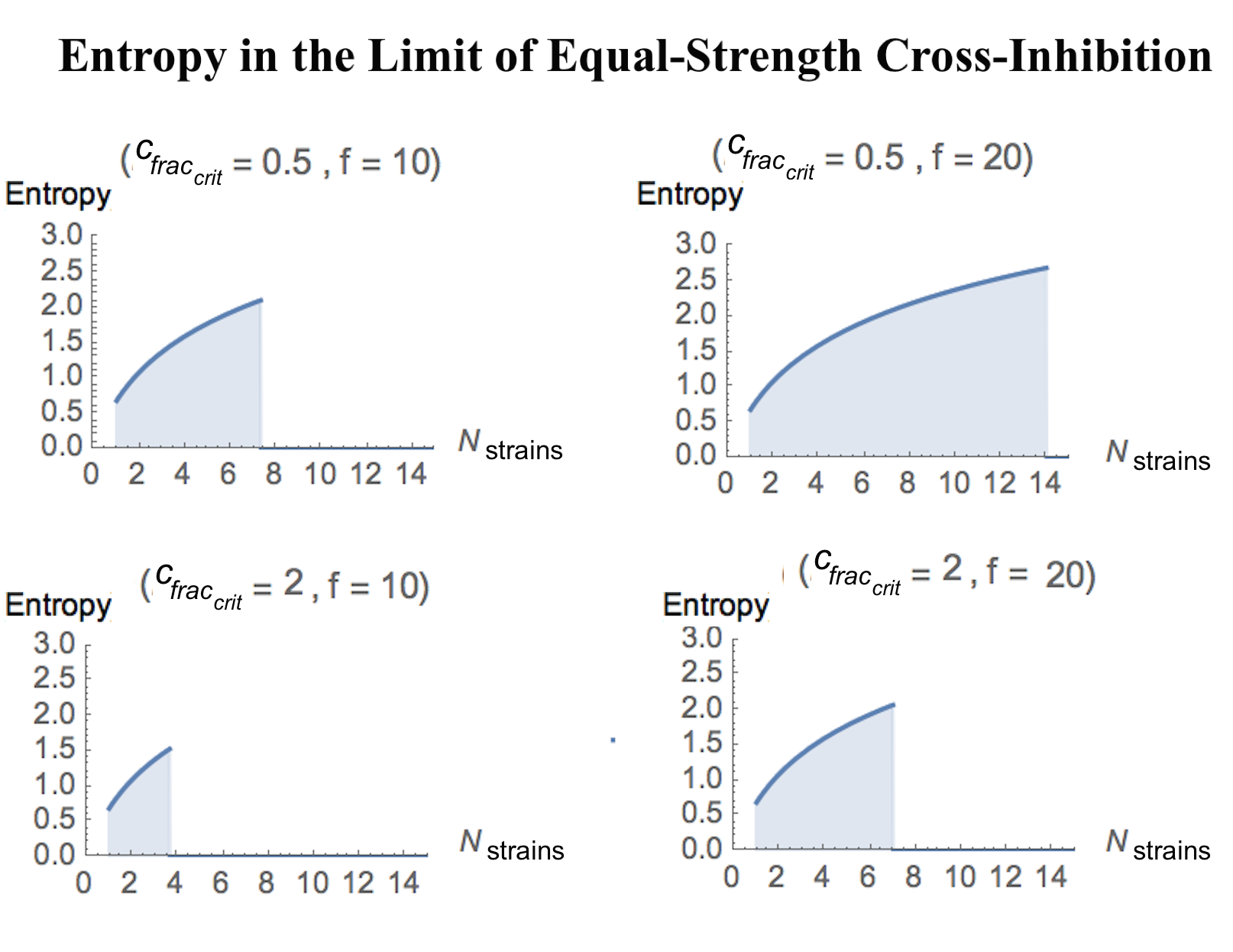}
\caption{In the limit of cross-group inhibition equal in strength to in-group activation, the capacity, as measured by the Boltzmann entropy, improves with strain number, up until a critical strain threshold $N_{strains_{c}}$. As discussed in the text, this critical value is determined by system parameters, and evaluates to $N_{strains_{c}} = (f+1)/(c_{frac_{crit}}+1)$}
\label{Fig:Capacity}
\end{figure}

\begin{figure}
\includegraphics[scale=.32]{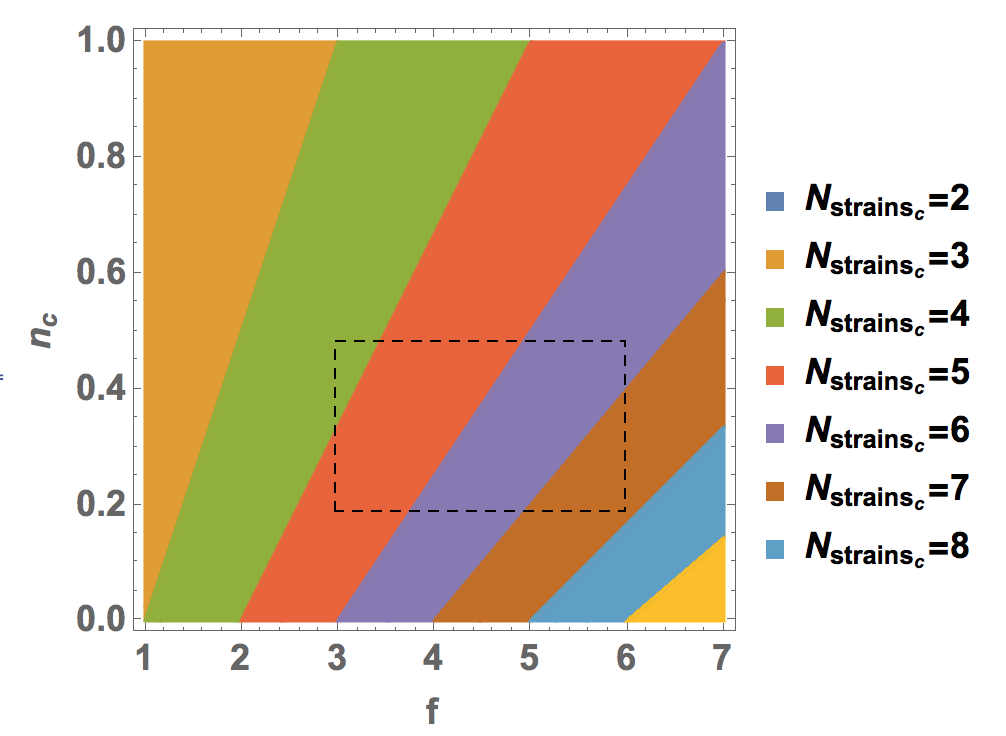}
\caption{The critical number of strains, at which point capacity is maximized just before dropping to zero, is regulated by the fold change $f$ and the critical cell density fraction $c_{frac_{crit}}$. The dashed rectangle indicates estimated parameter ranges for the \textit{Staphlococcus aureus} Agr two-component system.}
\label{Fig:NStrainsCrit}
\end{figure}

This result is plotted for various values of $f$ and $c_{frac_{crit}}$ in Figure \ref{Fig:Capacity}. As is seen there, the capacity increases with strain number, up until a threshold number of strains, 
\begin{equation}
N_{strains_{c}} = (f+1)/(c_{frac_{crit}}+1)
\label{eq:Nsc}
\end{equation}
at which point the increased amount of cross-inhibition due to the large number of differentiated strains completely suppresses the stability of any activated levels. Thus, if a quorum sensing community wants to maximize the number of different collective decisions that it can make, one possible strategy is to diversify into $N_{strains_{c}}$ different cross-inhibiting groups. The value of this optimal number of strains is dependent on both the fold change $f$ and the critical cell density fraction $c_{frac_{crit}}$, as illustrated in Figure \ref{Fig:NStrainsCrit}. 

\section{Application to \textit{Staphylococcus aureus}}
The model of equal-strength cross-inhibition is useful for analyzing the structure and function of the Agr two-component system, a quorum sensing circuit that regulates virulence in \textit{Staphylococcus aureus} \cite{Novick1995,Novick1997}. Four pherotypes, or genetic polymorphisms of the Agr locus, are observed in wild-type isolates. Strains within a specific pherotype activate members of their own pherotype, but inhibit members of different pherotypes, with an inhibitory strength that has been shown to be approximately equal to the cognate activation strength \cite{Mayville1999,Geisinger2009}. The evolutionary origins and significance of this diversifying selection remain incompletely understood. The neural-network-based formalism developed in this study allows for a fresh perspective on these questions.

\begin{figure}
\includegraphics[scale=.11]{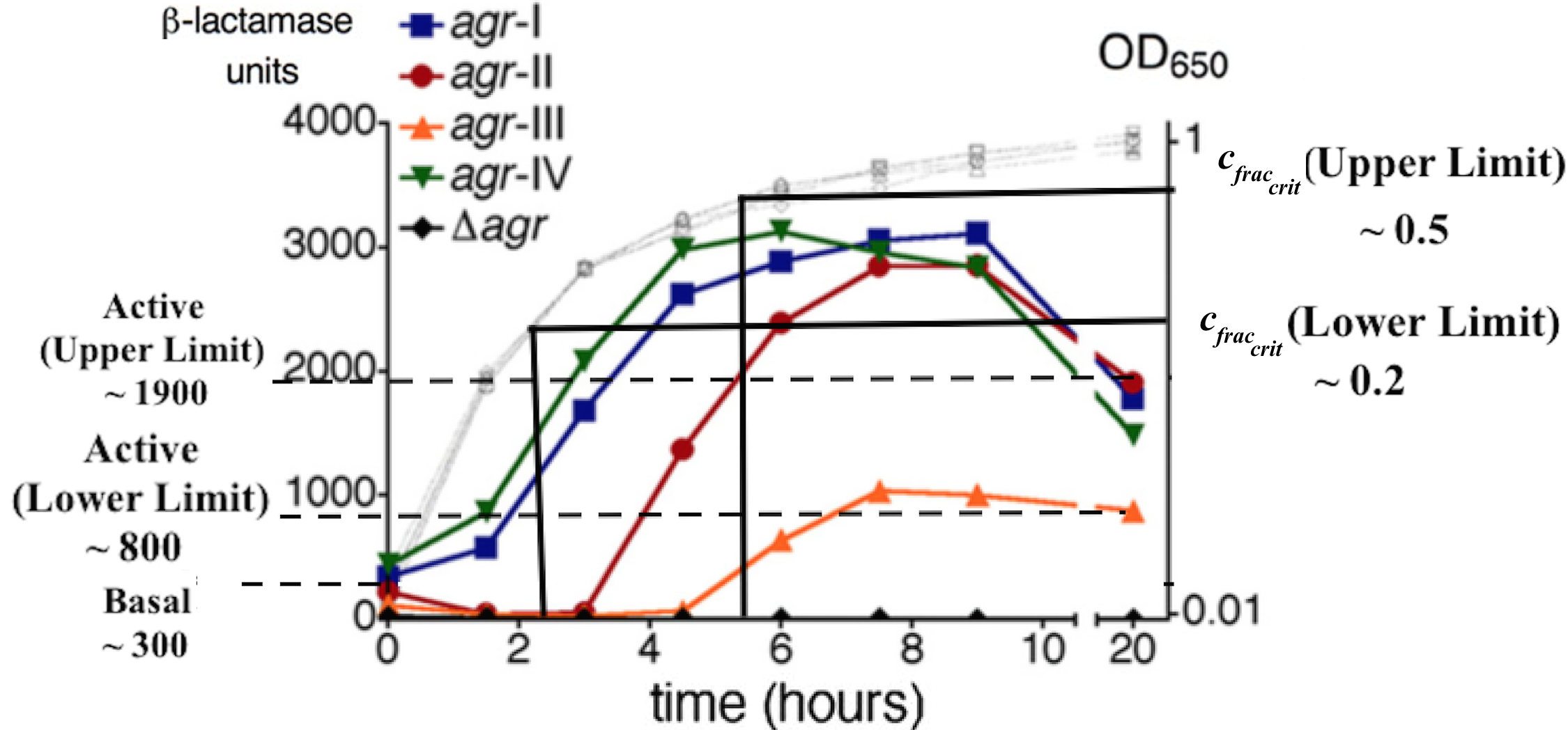}
\caption{Displayed is an adaptation of Figure 2(b) from Reference \cite{Geisinger2012nc}, presenting a comprehensive set of experimental measurements on allele-dependent variation in quorum sensing behavior (reprinted with permission from ASM). Observed is a characteristic range of fold changes $f$, as determined by the basal and maximal activity levels (dashed lines), and threshold cell density fractions $c_{frac_{crit}}$, as determined by the normalized optical densities of cell growth at which QS activity reaches half-maximal levels (solid lines).} 
\label{Fig:GeisingerData}
\end{figure}

Recent \textit{in vivo} and \textit{in vitro} work \cite{Geisinger2012nc} experimentally constrains the critical cell density density fraction $c_{frac_{crit}}$ to be between 0.2 and 0.5, and the fold change $f$ to be between 3 and 6, as shown in Figure \ref{Fig:GeisingerData}. Plugging this range of parameters into Equation (\ref{eq:Nsc}), we observe an expected range of $N_{strains_{c}}$ between approximately 2.5 to 6, precisely the range in which we find the actual observed number of pherotypes, $N = 4$. 

\subsection{Discussion and Outlook}
While far from conclusive, the observed correspondence between the actual number of pherotypes and the maximum-entropy prediction does inspire the question as to what possible fitness advantages, if any, are afforded by being near a critical point, where the information capacity reaches its maximum value prior to a sudden drop. This is especially relevant in light of mounting evidence in recent years that biological systems display self-organized criticality, naturally evolving towards a state that is an optimal balance of order and disorder \cite{Bialek2011,KrotovMorphogenesis,HidalgoInformationBasedFitness,Hidalgo2016}. In the context of intercellular communication networks, being near such a critical point would allow for collective group decision making that is simultaneously robust and flexible.

Although there has been much speculation on the existence of such elegant and beautiful unifying principles, more work is needed to bridge the gap between these abstract theoretical constructs and the actual biological networks found in nature. Here we have presented some initial steps towards bridging this gap. Although we have made somewhat artificial approximations of the actual biochemistry, it is straightforward, if tedious, to adapt the present formalism to incorporate more realistic features. For example, the assumption of a binary McCulloch-Pitts neuron can be relaxed to allow for graded sigmoidal \cite{DayanAbbott} activation dynamics that more accurately represents sigmoidal kinetics. Spatial structure can be incorporated by generalizing the well-mixed system of ordinary differential equations to a system of reaction-diffusion partial differential equations \cite{Ermentrout, YusufalyBoedicker2016}. Intrinsic noise and dynamical adaption of the input-output relations can be dealt with using standard tools from the statistical mechanics of learning and information \cite{StatMechLearning, StatMechInformation}. 

In summary, we have demonstrated a novel approach to compressing chemical details into simplified, generalizable neural network models, in order to study communication in complex real-world microbial ecosystems. The unexpected relationship between intercellular signaling dynamics and neural networks promises to be a valuable new theoretical tool for future studies of communication in heterogeneous microbial communities, with implications for both the basic science of microbial ecology and evolution, and also for synthetic biology efforts to engineer novel emergent behaviors in artificial multicellular consortia. 

In the immediate future, the formalism promises to be particularly valuable in efforts to characterize the biological function of crosstalk. In real ecosystems, one frequently finds a diverse combination of several coexisting biochemical languages, and the effects of the signals are, in general, non-additive. The gene regulatory behavior of the community under the influence of $N$ different signals simultaneously is not, in general, discernible from a knowledge of the behavior under each of the signals individually - the network exhibits additional `systems-level' properties that remain largely unexplored from the theoretical point of view. In large part, this has been due to the lack of sufficiently general, powerful methods for treating these complex effects. Based on the results presented in this study, we believe that a promising solution to this bottleneck may lie in importing concepts and tools from the field of neural networks, in order to rigorously and quantitatively characterize the functional role of information processing in social networks of microbes.

\section{Acknowledgments}
We gratefully acknowledge funding from DARPA YFA Grant No. D16AP00121 and useful discussions with all members of the Boedicker group, in particular Pavan Silva.
\section{Appendix}

\subsection{Microscopic Derivation of Equation (\ref{eq:effective})}
The derivation of Equation (\ref{eq:effective}) is copied almost directly from Chapter 29 of Reference \cite{KenDillBook}. We will present the argument for the case of two AI ligands competing for the same receptor binding site; generalization to arbitrary numbers is straightforward. For simplicity, we assume that the only biochemical difference between variant AIs and receptors is in the affinity for initial binding, and that any subsequent downstream steps are independent of the chemical nature of that binding; the arguments can be generalized to allow for differences in the downstream steps, but this is beyond the scope of the present work.

Suppose there are two AI variants $n_{1}$ and $n_{2}$, each binding to receptor $R_{1}$ with characteristic affinities:
\begin{subequations}
\begin{align}
R_{1} + n_{1} \overset{k_{11}}{\rightarrow} R_{1} n_{1} \\
R_{1} + n_{2} \overset{k_{12}}{\rightarrow} R_{1} n_{2}.
\label{eq:TwoAI}
\end{align} 
\end{subequations}
The binding site of receptor $R_{1}$ can be in three possible states: a) unbound, which we can take to have statistical weight 1, b) bound to $n_{1}$, with a statistical weight $k_{11} n_{1}$, by definition of the binding affinity, or c) bound to $n_{2}$ with a corresponding weight $k_{12} n_{2}$. 

In the limiting case where there is no competing AI, $n_{2} = 0$, it is straightforward to see that any biological `action' is proportional to the probability $p_{bound}$ that $R_{1}$ is bound to $n_{1}$. We can calculate this probability by dividing the weight of this binding event by the sum of all possible weights:
\begin{align}
\lim_{n_{2}\to0} \text{action} &\propto \lim_{n_{2}\to0} p_{bound} \\ 
&= \frac{k_{11} n_{1}}{1 + k_{11} n_{1}} \nonumber \\
&= \frac{1}{(k_{11} n_{1})^{-1} + 1} \nonumber
\end{align}

We can interpret $k_{11}^{-1}$ as a Michaelis-Menten constant - in the language of the main manuscript, $k_{11}^{-1} = n_{c}$, the concentration at which the system activates.

If the competing AI $n_{2}$ is present, the above expression must be generalized, but the form of the generalization depends on whether the competing AI is an activator or an inhibitor. We present each case separately, as the physical arguments for each scenario differ slightly.

\subsubsection{Competitive Activation}
If $n_{2}$ is an activator, then there is still biological action when it binds to $R_{1}$, and thus, the only thing that changes is that the binding probability calculation contains extra terms corresponding to the additional statistical weight:
\begin{align}
\text{action}_{activation} &\propto p_{bound_{activation}} \\ 
&= \frac{k_{11} n_{1} + k_{12} n_{2}}{1 + k_{11} n_{1} + k_{12} n_{2} } \nonumber \\
&= \frac{1}{(k_{11} (n_{1} + \frac{k_{12} n_{2}}{k_{11}}))^{-1} + 1} \nonumber .
\end{align}

From this we see that the kinetics of the system in the presence of a competitive activating AI is formally equivalent to that in the absence of the AI, with in an effective shift of the cognate AI concentration $n_{1}$,
\begin{equation}
n^{eff}_{1} = n_{1} + \frac{k_{12}}{k_{11}} n_{2} = n_{1} + w_{12} n_{2}
\end{equation}
where we have defined the weight $w_{12} = \frac{k_{12}}{k_{11}} > 0$.

\subsubsection{Competitive Inhibition}
If instead $n_{2}$ is an inhibitor, now any biological action is restricted to instances where $R_{1}$ binds to $n_{1}$, and only $n_{1}$. Then, in the calculation of the `active' binding probability, while the denominator still contains weight terms for three possible scenarios, now the numerator only includes one statistical weight,
\begin{align}
\text{action}_{inhibition} &\propto p_{bound_{inhibition}} \\ 
&= \frac{k_{11} n_{1}}{1 + k_{12} n_{2} + k_{11} n_{1} } \nonumber \\
&= \frac{1}{ (\frac{k_{11}}{1 + k_{12} n_{2}} n_{1})^{-1} + 1} \nonumber
\end{align}
Here, the kinetics in the presence of a competitive inhibiting AI is formally equivalent to that in the absence of the AI, if we effectively shift $n_{c} = k_{11}^{-1}$,
\begin{equation}
n^{eff}_{c} = (k^{eff}_{11})^{-1} = k_{11}^{-1} + \frac{k_{12}}{k_{11}} n_{2} = n_{c} + \frac{k_{12}}{k_{11}} n_{2} .
\end{equation}
In the limit of a step-function approximation for the nonlinear activation kinetics, as done in the main text, this shift of $n_{c}$, with a fixed value of $n_{1}$, is equivalent to a fixed value of $n_{c}$, with a shift instead in $n_{1}$:
\begin{equation}
H(n_{1} - n^{eff}_{c}) = H(n_{1} - \frac{k_{12}}{k_{11}} n_{2} - n_{c}) = H(n^{eff}_{1} - n_{c})
\end{equation}
with
\begin{equation}
n^{eff}_{1} = n_{1} - \frac{k_{12}}{k_{11}} n_{2} = n_{1} + w_{12} n_{2}
\end{equation}
where the weight $w_{12} = - \frac{k_{12}}{k_{11}} < 0$ is now negative due to the interaction being inhibitory.

\subsection{Robustness Analysis}
\begin{figure}
\includegraphics[scale=0.36]{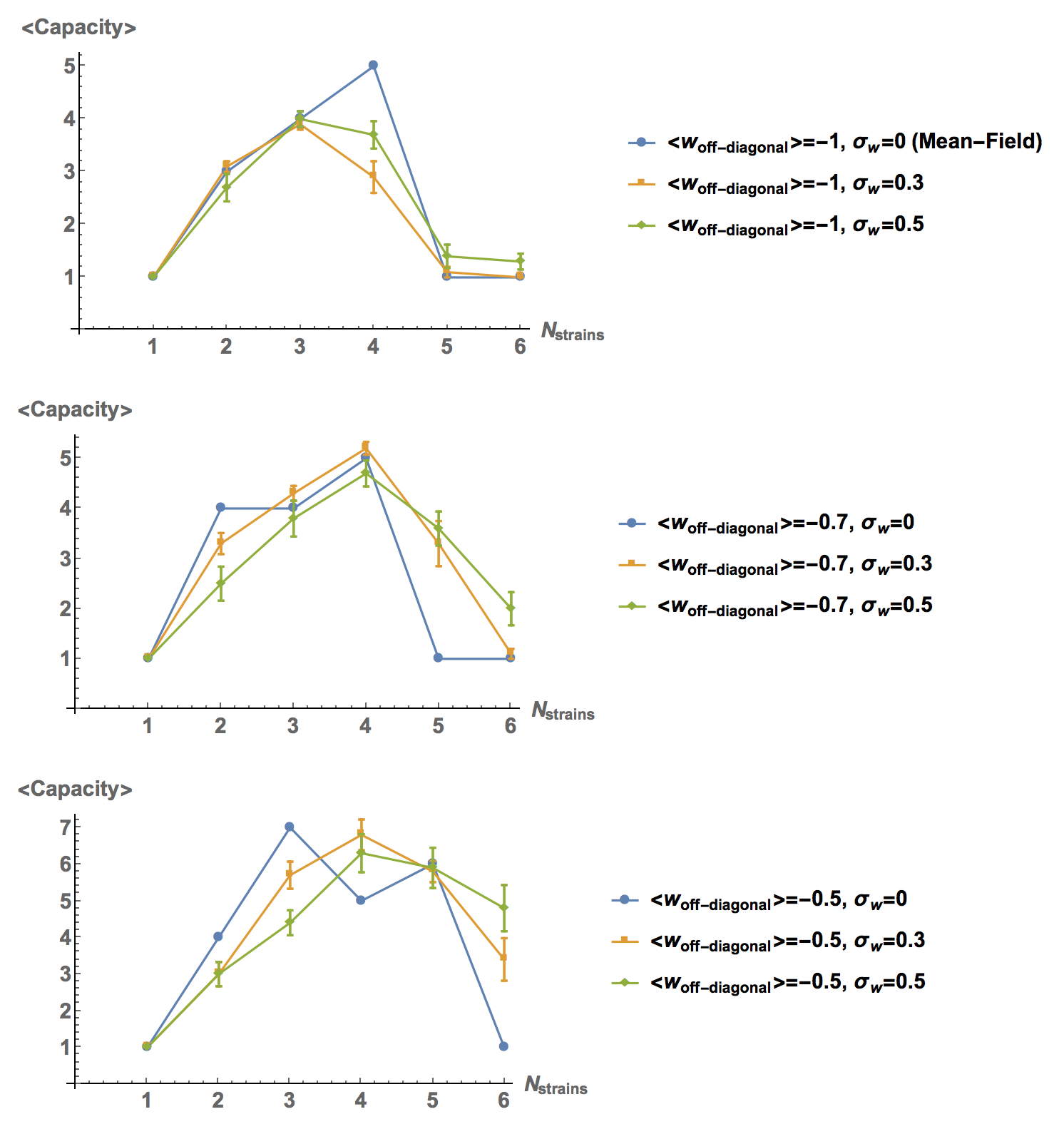}
\caption{The qualitative scaling of the capacity with strain number for various heterogeneous parameter distributions is found to be well approximated by the mean-field model with $\langle w_{off-diagonal} \rangle = -1$. For each trial, we run ten replicates, and take the mean and standard error of the capacity to be the data value and uncertainty, respectively.}
\label{RobustnessWeightsPlot}
\end{figure}

In the main manuscript, we made a `mean-field' approximation to simplify the system and allow for exact analytical treatment. Here, we will show that several of the qualitative results obtained with this approximation carry over to some more general cases. The essence of the mean-field approximation, in a nutshell, can be summarized by the statement that the probability distribution of the off diagonal weights, $w_{off-diagonal}$ was a Dirac Delta function centered at -1:
\begin{align}
p(w_{off-diagonal}) &\approx p^{MFT}(w_{off-diagonal}) \\
&= \delta(w_{off-diagonal} + 1) \nonumber .
\end{align}

We may generalize this to account for more realistic weight distributions by: 1) allowing the average off-diagonal affinity to not be negative unity, $\langle w_{off-diagonal} \rangle \neq -1$, and 2) making $w_{off-diagonal}$ a normally distributed variable with variance $\sigma_{w}$:
\begin{align}
p(w_{off-diagonal}) &= \\
& \frac{1}{\sqrt{2 \pi \sigma_{w}^{2}}} e^{-\frac{(w_{off-diagonal} - \langle w_{off-diagonal} \rangle)^{2})}{2 \sigma_{w}^{2}}} \nonumber .
\end{align}

Then, instead of calculating a single Boltzmann entropy as a function of strain number $S(N)$ (or, equivalently, capacity as measured by $e^{S(N)}$), we must instead calculate the average $\langle S(N) \rangle$ (or $\langle \text{capacity} \rangle$) over many random samples from this weight distribution. 

In Figure \ref{RobustnessWeightsPlot}, we plot the capacity, as measured by $e^{S(N)}$, for several values of $\langle w_{off-diagonal} \rangle$ and $\sigma_{w}$, using the average $S. \ aureus$ parameter values (as shown in the main text) of $f = 4.5$ and $c_{frac_{crit}} = 0.35$. There we see that the qualitative prediction of the capacity slowly increasing up to an `optimal' range of strain variants, prior to rapidly dropping beyond a `tipping point', holds even for these more realistic weight distributions. The precise shape of the curve and slope of the downfall beyond the tipping point are modified, but the local maximum remains fairly strongly localized around $N_{strains} = 4$. Future work should look into developing more sophisticated analytical approaches to explain these numerical results and more precisely demarcate the regime of validity of the mean-field approximation.

\end{document}